# Successive magnetic phase transitions with magnetoelastic and magnetodielectric coupling in the ordered triple perovskite $Sr_3CaRu_2O_9$

Arun Kumar[1], Pascal Manuel[2] and Sunil Nair[1]

[1]Department of Physics, Indian Institute of Science Education and Research Dr. Homi Bhabha Road, Pune, Maharashtra-411008, India

[2]ISIS Pulsed Neutron Source, STFC Rutherford Appleton Laboratory, Didcot, Oxfordshire OX11 0QX, United Kingdom

**Abstract:**

We report a comprehensive temperature-dependent investigation of the 1:2 ordered triple perovskite system $Sr_3CaRu_2O_9$. It crystallizes in the monoclinic structure with space group $P2_1/c$, consisting of corner-sharing $CaO_6$ and $RuO_6$ octahedra. Using DC magnetization and neutron diffraction measurements, we show that this system undergoes successive magnetic transitions ~190 K and ~160 K. From the analysis of the temperature-dependent neutron diffraction and dielectric data, we demonstrate two distinguishing features of the $Sr_3CaRu_2O_9$ system: (i) magnetoelastic coupling associated with the two magnetic transitions, as revealed by a change in the unit cell volume, direct Ru-Ru distance, and bond angles and (ii) magnetodielectric coupling, as revealed by the anomalies across the two magnetic transitions. The present results would add significantly to the current understanding of the triple perovskites with incipient spin-orbit coupling.

**I. Introduction:**

The 3d transition metal-based perovskite oxides have continued to occupy a prominent place in the area of strongly correlated systems. In the recent past, the study of hybrid 3d - 4d/5d transition metal oxide-based triple perovskites with the general formula $A_3BB'_2O_9$ (where A = Ba, Sr and B = alkali metals, alkaline earth metals, or lanthanides, B' = Ru, Ir, Os, Nb, Ta, Sb) has become a very active area of research due to their wide-ranging functional properties with potential for technological applications [1-13]. The intricate interplay between the strong spin-orbit coupling, electronic correlations, crystal field effects and Hund's coupling in such perovskites [1] lead to novel electronic and magnetic properties such as Mott insulators [2], quantum spin-liquids [3–5], quantum spin-orbital liquid [6], spin-glass [7], long-range antiferromagnetic (AFM) [8,9] /ferromagnetic (FM) ordering [10,11] and multiferroicity [12,13].

Within the triple perovskites, Ba-based triple perovskite ruthenates and iridates of the form $Ba_3BB'_2O_9$ (B = alkali metals, alkaline earth metals, or lanthanides and B'= Ru, Ir) have been widely investigated for their structure-property correlations [9–11,14–19]. Most derivatives usually adopt the aristotype *6H*-hexagonal structure in which Ba ion occupies the 12-fold coordinated site while B and B' occupy the corner and face-shared sites, respectively, forming $BO_6$ octahedra and $B'_2O_9$ dimers comprising of two face-shared $B'O_6$ octahedra. Interestingly, the $B'_2O_9$ dimer units arranged along the c-axis produce a triangular lattice of B' atoms in the a-b plane. Based on such structural features and choice of non-magnetic or magnetic *B*-cations, these triple perovskite oxides exhibit diverse structural as well as magnetic phenomena such as spin-dimer [14], spin-gap opening [15,16], charge ordering [17], long-range AFM ordering [9–11], magneto-structural transitions [9,18], magneto-elastic coupling [16] and magnetodielectric coupling [19,20].

In contrast, the investigation of Sr-based triple perovskites of the form $Sr_3BB'_2O_9$ has been limited to a few systems [5,8,21,22]. These Sr-analogues are characterized by monoclinic symmetry in which B and B' cations form corner-sharing $BO_6$ and $B'O_6$ octahedra and feature a unique buckled honeycomb lattice of B' atoms. Among the family of Sr-based triple perovskites, $Sr_3CaRu_2O_9$ is particularly interesting, as it was the first compound to exhibit a 1:2 cation ordering at the *B*-site in transition metals with partially filled d-orbitals [23] in contrast to earlier reported $d^0$-cations (non-magnetic) like $Nb^{5+}$, $Ta^{5+}$ [24]. To the best of our knowledge, no detailed (magnetic, dielectric or transport) investigations were carried out on $Sr_3CaRu_2O_9$, except the room temperature crystal structure [23]. It is of interest to investigate $Sr_3CaRu_2O_9$ to better understand the structure-property correlations in the family of triple perovskite ruthenates. The two Sr-based 5d triple perovskites $Sr_3CaIr_2O_9$ and $Sr_3CaOs_2O_9$ have been investigated in the past for their room-temperature crystal structure and magnetic properties [8,21]. Both are reported to crystallize in the monoclinic structure in the *$P2_1/c$* space group but exhibit different magnetic behaviour [8,21]. The $Sr_3CaIr_2O_9$ shows paramagnetic-like behaviour below room temperature [21], while $Sr_3CaOs_2O_9$ exhibits long-range ordered antiferromagnetic (AFM) transition with *$T_N$* ~385 K [8] which is the highest among all the members of the triple perovskites family. Theoretically, it has been argued that the sizeable antiferromagnetic exchange interaction between the puckered planes results in a high *$T_N$* [8].

Here, we report on $Sr_3CaRu_2O_9$, a hitherto unexplored member of the triple perovskite family, using a combination of x-ray diffraction, neutron powder diffraction, dc magnetization, specific heat, dielectric and resistivity measurements. It is shown that $Sr_3CaRu_2O_9$ crystallizes in the monoclinic symmetry with *$P2_1/c$* space group. Temperature-dependent dc magnetization and neutron diffraction analysis reveal two magnetic transitions ~160 K and ~190 K below room temperature. Further, the temperature dependent dielectric permittivity also exhibits anomalies across the two magnetic transitions suggesting the possibility of magneto-dielectric coupling.

## II. Experimental details:

Single-phase polycrystalline specimens of $Sr_3CaRu_2O_9$ (SCRO) were synthesized by the conventional solid-state method using high-purity carbonates $SrCO_3$ ($\geq$ 99.999 %, Sigma Aldrich), $CaCO_3$ ($\geq$ 99.95 %, Sigma Aldrich) and oxide $RuO_2$ ($\geq$ 99.9 %, Sigma Aldrich) as starting materials. Stoichiometric amounts of precursor powders were thoroughly ground for 6 hrs in an agate mortar and pestle with ethanol as a mixing medium. The homogeneously mixed powder was first calcined at 1123 K for 12 hrs in air - followed by re-calcination at 1473 K twice for 48 hrs each with intervening regrinding. Subsequently, the calcined powder was pressed into pellets (13 mm diameter and 1.0 mm thickness) and sintered at 1498 K for 12 hrs. The pellets were crushed into a fine powder and then annealed at 773 K for 12 hrs to remove the strain produced during crushing. The well-annealed powder was used for the x-ray diffraction and neutron scattering measurements.

The elemental compositions were checked by the energy dispersive x-ray (EDS) spectroscopic technique attached with field emission scanning electron microscope (Model-Zeiss Ultra Plus). X-ray diffraction (XRD) measurements were performed using a high-resolution powder diffractometer operated in the Bragg-Brentano geometry (Bruker, Model-D8 ADVANCE). Temperature and field-dependent dc magnetization measurements were carried out using a superconducting quantum interference device (SQUID)-based magnetometer (MPMS-XL, Quantum Design, USA). Temperature-dependent heat capacity was measured by the relaxation method using a physical property measurement system (PPMS, Quantum Design, USA). Temperature-dependent resistivity measurements were carried out using the four-probe method in the same PPMS. Temperature-dependent dielectric measurements were performed using a close cycle refrigerator (CCR) system and a Novocontrol (Model-Alpha-A) high-performance frequency analyzer. Temperature-dependent time-of-flight neutron diffraction experiments were performed using the WISH diffractometer at ISIS Facility, UK [25]. The XRD and neutron diffraction data were analyzed by the Rietveld refinement technique using FullProf software [26]. The crystal and magnetic structures were drawn using the VESTA software [27].

**III. Results and discussion:**

The average atomic percentage ratio determined by the EDS analysis Sr: Ca: Ru ≈ (2.93 ± 0.05): (0.99 ± 0.05): (2.10 ± 0.06) are very close to the nominal compositions - confirming the excellent quality of our sample. The crystal structure of SCRO was analysed by laboratory x-ray powder diffraction (XRD), and the room temperature XRD pattern is shown in Fig. 1(a). Analysis of the XRD pattern confirms the single-phase formation without any impurity phase. All the peaks could be indexed with the monoclinic phase in the $P2_1/c$ space group, as confirmed by the Rietveld refinement technique. The supercell of the SCRO consists of 16 independent atoms, all of which occupy the general positions except for two Ca-ions which are fixed at the special positions. In the refinements, the background was modelled using linear interpolation between the data points. The peak shape was modelled using a pseudo-Voigt function. The Full-Width at Half-Maxima (FWHM) of the peaks were modelled using the Caglioti equation [28] $(FWHM)^2 = U \tan^2\theta + V \tan\theta + W$. During the refinement, scale factor, zero displacement, lattice parameters (a, b, c), atomic coordinates (x, y, z) and thermal parameters (B) were allowed to vary. Fig. 1(a) depicts the observed (red-filled circles) and calculated (black continuous line) XRD profiles of SCRO for the monoclinic $P2_1/c$ space group. Both profiles show excellent fit as can be seen from the nearly smooth difference profile shown in the bottom line (blue continuous line) of the same figure. Table 1 summarizes the structural parameters of SCRO after the Rietveld refinement using $P2_1/c$ space group which agree well with the values reported in the literature [23]. A schematic crystal structure is shown in Fig. 1(b), which can be described as 1:2 ordering of $Ca^{2+}$ and $Ru^{5+}$ cations over the six coordinated *B*-site of the perovskite where the $CaO_6$ octahedra are corner shared with $RuO_6$ octahedra and $Sr^{2+}$ cations occupying the 12-fold coordinated A-site. It can also be visualized as a layered structure with stacking sequence of (one layer of) $CaO_6$ octahedra and (two layers of) $RuO_6$ octahedra along the crystallographic b-axis. The 1:2 cation ordering is driven by significant charge and size difference between $Ca^{2+}$ (1.00 Å) and $Ru^{5+}$ (0.565 Å) cations. The important bond-lengths are listed in Table 2. It is evident from the table that the Ru-O bond-length lies in the range of 1.82-2.10 Å while Ca-O bond-length are in the range 2.20-2.24 Å. As a result of the difference in the bond lengths, the $RuO_6$ and $CaO_6$ octahedra are highly distorted. The average Ru-O bond length (~1.96 Å) is very close to that anticipated from sum of the ionic radii [29], while the average Ca-O bond-length (~2.21 Å) appears to be slightly shorter than expected. The Ru-O-Ru bond angles between the nearest-neighbour deviate significantly from the ideal 180° due to the octahedral tilting. The smallest Ru-Ru distance ~3.92 Å is formed via Ru-O bonds of very similar lengths (2.01 Å and 1.96 Å) while the longest Ru-Ru distance ~ 4.00 Å is formed via Ru-O bonds of significantly different lengths (2.10 Å and 1.94 Å). This Ru-Ru distance (~ 4.00 Å) is much larger than that observed in the $Ba_3CaRu_2O_9$ (~2.6-2.7 Å) which usually forms a $Ru_2O_9$ dimer of two face-sharing octahedra [14]. The corner-connected tilted $RuO_6$ octahedra form a unique buckled honeycomb lattice of Ru, where the distance between the two adjacent Ru atoms is ~ 4.00 Å. Such a buckled honeycomb lattice has also been observed in the related isostructural triple perovskites $Sr_3CaIr_2O_9$ and $Sr_3CaOs_2O_9$ [8,21].

Temperature dependence of zero-field cooled (ZFC) dc susceptibility $\chi$ *(T)* measured under the application of 1000 Oe field is depicted in Fig. 2 (a). It exhibits an anomaly at Néel temperature $T_N$ ~ 190 K associated with the long-range antiferromagnetic ordering of the Ru spins. This is further confirmed by our neutron diffraction studies. Below $T_N$, $\chi$ *(T)* decreases with decreasing temperature and shows a smeared broad anomaly ~160 K. On further decreasing the temperature, $\chi$ *(T)* shows an upturn below ~ 70 K and increases continuously down to the lowest temperature of measurements. A similar upturn has also been observed in the $Ba_3CaRu_2O_9$, $Sr_3CaOs_2O_9$ and attributed to the presence of trace amount of paramagnetic impurities [8]. Above $T_N$, the $\chi$ *(T)* is paramagnetic but it does not obey a Curie-Weiss behaviour. Consequently, it is not possible to estimate the strength of interaction i.e., Curie-Weiss temperature, and effective magnetic moment. It is worth comparing the observed magnetic behaviour of SCRO to other isostructural triple perovskites $Sr_3CaIr_2O_9$, $Sr_3CaOs_2O_9$. Despite having the same type of octahedral sharing of magnetic ions and similar bond-lengths, bond-angles, and direct Ir-Ir, Os-Os distances, the susceptibility of $Sr_3CaIr_2O_9$ displays a paramagnetic-like behaviour without any anomaly in the temperature range 2-300 K [21] whereas $Sr_3CaOs_2O_9$ is found to be ordered antiferromagnetic below $T_N$ ~385 K [8]. Intriguingly, the occurrence of two magnetic transitions have also been observed in the family of Ba-based triple perovskite systems like $Ba_3CoNb_2O_9$ [30], $Ba_3CoSb_2O_9$ [31], $Ba_3CoTa_2O_9$ [32] and $Ba_3MnNb_2O_9$ [13] with a single magnetic ions at the *B*-site. Notably, the magnetic transition temperatures of the above reported systems are much lower than that observed in our SCRO system. The magnetic ordering in SCRO is further investigated by the specific heat ($C_p$) measurement, which is shown in panel (b) of Fig. 2. It is evident that the temperature dependence of $C_p(T)$ exhibits a lambda ($\lambda$)-like anomaly at $T_N$ ~190 K. This again confirms the onset of long-range antiferromagnetic ordering consistent with dc susceptibility results.

To get further insights into the nature of the observed anomalies, we plot the temperature dependence of real-part of dielectric permittivity ($\varepsilon'$) at selected frequency of 901 kHz in Fig. 2(c). It has been reported that at such high frequencies (> 100 kHz), the extrinsic contribution to the dielectric permittivity is negligible [33]. We note that $\varepsilon'$ increases monotonically with increasing temperature but exhibit two anomalies across the two magnetic transition. These anomalies are more clearly shown in the insets of Fig. 2(c). The temperature dependence of tanδ exhibits a peak which coincides with the long-range ordering transition while a broad peak feature is also observed at the second magnetic transition of 160 K in the real part of ε'. The appearance of an anomaly in the ε' /tanδ across a magnetic transition is suggestive of magneto-dielectric coupling [20]. We believe that our results open a new avenue to explore the magneto-dielectric effects in the triple perovskite family which is scarcely investigated.

To get a microscopic understanding of the low-temperature magnetic transitions of SCRO, we carried out neutron powder diffraction (NPD) studies in the 250 to 1.5 K temperature range. Fig. 3(a) displays the temperature-dependent evolution of NPD patterns over a selected range of ToF ~17-85 ms (scattering angle = 58.3 deg and flight path length = 42.2 m) for SCRO. It is evident that no new peaks appear or disappear in the investigated temperature range 1.5 - 250 K. This implies that the crystal structure remains monoclinic in the $P2_1/c$ space group and that there is no structural phase transition in this temperature range. Moreover, around T ≈ 200 K, we observed that the two Bragg peaks (~ 51.5 mToF and 27.5 mToF) start showing additional intensity contributions, which are due to the magnetic ordering of the Ru spins. The temperature evolution of the strongest magnetic peak after subtracting the 220 K pattern as a paramagnetic background is shown in Fig. 3(b), which clearly confirms the appearance of long-range ordered transition at $T_N$ ~200 K. This is in close agreement with $T_N$ observed in the dc susceptibility and specific heat measurements.

The temperature variation of the integrated intensity of the strongest magnetic peak ($I_{mag}$), containing two overlapping Bragg reflections (-111) and (-202), is shown in Fig. 3(c). Evidently, $I_{mag}$ decreases with increasing temperature and completely disappears ~ 200 K. We have modelled the $I_{mag}$ using a power law over a limited temperature range near $T_N$ (0.596 ≤ $T/T_N \leq 1$) [34]: $I(T) \propto (1-T/T_N)^{2\beta}$, where $T_N$ and $\beta$ represent the Néel temperature and critical exponent, respectively. The solid line through the data points is the best fit with $T_N = (201.3 \pm 0.2)$ K and $\beta = (0.232 \pm 0.006)$, respectively. It is interesting to note that the observed value of the critical exponent $\beta$ is very close to that reported for a tricritical phase transition ($\beta = 0.25$) [35]. A similar exponent has also been reported in compounds like $Cu_3Nb_2O_8$ [34], $(Ba_{1-x}Bi_x)(Ti_{1-x}Fe_x)O_3$ [36] and $EuAl_2Ge_2$ [37].

The magnetic structure compatible with the monoclinic $P2_1/c$ space group symmetry is determined by representation analysis using the BasIreps program implemented in the Fullprof suite [26]. The magnetic contribution superimposed on the nuclear Bragg contribution suggests the propagation vector **k** = (0 0 0). In the $P2_1/c$ space group with a monoclinic unit cell, the two magnetic ions ($Ru_1^{5+}$ and $Ru_2^{5+}$) occupying the same 4e Wyckoff site with different general positions ($x_1$, $y_2$, $z_2$) and ($x_1$, $y_2$, $z_2$). The reducible magnetic representation $\Gamma$ can be decomposed in terms of four irreducible representations $\Gamma_k$ for the 4e site as follows:

$$\Gamma(4e/Ru^{5+}) = 3\Gamma_1^1 + 3\Gamma_2^1 + 3\Gamma_3^1 + 3\Gamma_4^1 \qquad \ldots\ldots (1)$$

where $\Gamma_1$, $\Gamma_2$, $\Gamma_3$ and $\Gamma_4$ are one-dimensional in nature. Each irreps contains three basis vectors and, therefore, three refinable parameters.

We observed only two magnetic Bragg peaks in the NPD patterns of SCRO which is not sufficient to unambiguously solve the magnetic structure. In order to determine the magnetic structure and ordered moment of the $Ru^{5+}$, we have made an assumption that the $Ru^{5+}$ ions on both sites have same magnetic moment. We have tested all four irreps against our NPD data. Out of four irreducible representations, $\Gamma_1$ and $\Gamma_3$ represent the ferromagnetic ordering with parallel arrangement of $Ru^{5+}$ spins and must be excluded according to susceptibility data. This leaves $\Gamma_2$ and $\Gamma_4$ as the likely representations for SCRO. The two representations can be distinguished by our Rietveld refinement analysis. The refinement corresponding to $\Gamma_2$ model reproduce the observed magnetic peak and gave the best fit with the lowest $R_{Mag}$ = 5.6 % at 1.5 K. The value of $R_{Mag}$ for $\Gamma_4$ representation is nearly seven times larger than that of the $\Gamma_2$ model. The simultaneous refinement using $\Gamma_2$ model yields the magnetic moment $mRu_1$ = (1.77 ± 0.04) $\mu_B$ and $mRu_2$ = -(1.77 ± 0.04) $\mu_B$ at 1.5 K which is much smaller than the expected spin-only value of 3 $\mu_B$ for $Ru^{5+}$ (S = 3/2). The observed value of the ordered moment is consistent with the many other reported Ru-and Os-based triple perovskite systems like $Sr_3CaOs_2O_9$ (1.88 (1) $\mu_B$) [8], $Ba_3LaRu_2O_9$ (1.4 (1) $\mu_B$) [10], $Ba_3NiRu_2O_9$ (1.5 (2) $\mu_B$) [38,39], $Ba_3CoRu_2O_9$ (1.44 (5)$\mu_B$) [38]. The strong suppression of the ordered moment in Ru-based systems has been attributed to a substantial hybridization effect between the Ru 4d and O 2p states [40]. Figs. 4 (a) and (b) depict the observed (filled-circles), calculated (continuous line) and difference profiles obtained after Rietveld analysis of NPD data at 250 K (paramagnetic state) and 1.5 K (magnetically ordered state), respectively. The observed and calculated profiles show satisfactory fit as can be seen from the nearly flat difference profile. The corresponding magnetic structure of SCRO using $\Gamma_2$ model is shown in Fig. 4 (c). The spins in the neighbouring sites are coupled antiferromagnetically and lie in the ac plane. The $Ru_1$ and $Ru_2$ spins are shown with black and blue colours, respectively.

Even though there is no structural phase transition in SCRO down to 1.5 K, the temperature dependence of the monoclinic unit cell volume (see Fig. S1 of the supplemental information [41] for temperature variation of lattice parameters), as obtained from the Rietveld refinement analysis of NPD data shown in Fig. 5 (a) reveals that the unit cell volume decreases linearly with decreasing temperature and shows a clear change of slope around 200 K followed by a smeared feature at the second transition ~160 K. The observation of an anomaly in the unit cell volume around the two magnetic transitions suggests magneto-elastic coupling [42]. Further, the temperature dependence of the direct Ru-Ru distance and some selected Ru-O-Ru bond angles are depicted in Figs. 5 (b) and 5 (c). Note that the both Ru-Ru distances are constant with decreasing temperatures upto the $T_N$ ~200 K. On further decreasing temperature one of the Ru-Ru distance increases and other decreases gradually upto the second magnetic transition ~160 K and then constant down to the lowest temperature. The bond angles also change significantly across the $T_N$ and increase/decrease upto 160 K. We believe that as a consequence of the monoclinic distortion, the multiple Ru-O-Ru bond angles (and bond lengths) are formed and thus creating multiple exchange interaction pathways that influence the magnetic properties of SCRO [11]. Thus our NPD analysis clearly reveals a significant change in the unit cell volume around the two transitions suggesting magnetoelastic coupling.

The temperature dependence of the electrical resistivity ($\rho$) of SCRO at zero magnetic fields is shown in Fig. 6. It is evident that the $\rho$ (T) increases with decreasing temperature indicating the insulating nature of the SCRO. Below 116 K, $\rho$ (T) goes beyond the measurement limit of the PPMS instrument. The value of resistivity increased several orders of magnitude at 116 K ($\rho$ ~ 4x10$^5$ ohm-cm) as compared to the room temperature resistivity ($\rho$ ~ 81 ohm-cm). To understand the transport mechanism, we attempted to fit the data to the Arrhenius law described by:

$$\rho = \rho_0 \exp(E_a/k_BT) \qquad (2)$$

where $\rho_0$ is a constant, $E_a$ is the activation energy and $k_B$ is the Boltzmann constant. As per Arrhenius Eq. (2), the ln ($\rho$) versus 1/T plot should be linear. The Arrhenius plot shown in the inset (i) of Fig. 6 exhibits linear behaviour only in the limited range 225-300 K and deviates very significantly at low temperatures which possibly suggests a 3D variable range hopping (VRH) mechanism [43]. The activation energy estimated from the fitting of the linear region comes out to be $E_a$ ~ 0.189 eV. This is in close agreement with the $E_a$ determined from the temperature-dependent dielectric study. Further, we attempted to fit Mott's 3D VRH model described by [43]:

$$\rho(T) = \rho_0 [\exp(T_0/T)^{1/4}] \qquad (3)$$

where $T_0$ is the Mott's activation energy (in units of temperature) which depends on the density of states at the Fermi level ($N(E_F)$) and localization length ($\xi$) through a relation ($T_0 \propto 1/N(E_F)\xi^3$). The linear nature of the ln ($\rho$) versus $T^{-1/4}$ plot in almost the entire temperature range (see the inset (ii) of Fig. 6) suggest that the transport mechanism of charge carriers in SCRO is well described by the 3D VRH model. This is similar to what is reported in the double perovskites $Sr_2CoRuO_6$ [44], $Ca_{2-x}Sr_xFeRuO_6$ [45] and triple perovskites $Sr_3CaOs_2O_9$ [8], $Sr_3CaIr_2O_9$ [21] and $Ba_3NiIr_2O_9$ [3]. The observed value of $T_0 = 3.7 \times 10^8$ K is in line with reported values for other Ru-based oxides [44,46].

Figure 7 presents the temperature dependence of the real part of dielectric permittivity ($\varepsilon'$) and loss tangent (tan$\delta$) at several frequencies ranging from 100 kHz – 901 kHz. It is notable that the $Sr_3CaRu_2O_9$ exhibits a large $\varepsilon'$ at room temperature ($\varepsilon' \sim 8600$ at 100 kHz) similar to that reported in the well-known relaxor ferroelectric-like materials $Pb(Mg_{1/3}Nb_{2/3})O_3$ [47], $Pb(Mg_{1/3}Nb_{2/3})O_3$-$PbTiO_3$ [48]. We observe two distinct frequency dispersions near the magnetic transitions in the $\varepsilon'$-T plot. These two frequency dispersions hereafter labelled as low-temperature (LT) and high-temperature (HT), suggest the presence of two dielectric relaxations in SCRO. In addition, the tan $\delta$ exhibits a peak at the inflection point of the $\varepsilon'$ at a particular frequency for the HT relaxation. We note that the peak position of tan $\delta$ shifts to higher temperature side with increasing frequency. The first derivative of $\varepsilon'$ (i.e., $d\varepsilon'/dT$) was used to determine the accurate peak position for the analysis of LT relaxation while the HT relaxation is analysed by the tan $\delta$ peak at the inflection point of $\varepsilon'$. The relaxation time ($\tau = 1/2\pi f$) corresponding to both peak temperatures follow simple Arrhenius behaviour ($\tau = \tau_0 \exp(E_a/k_BT)$), as can be seen from the linear nature of the ln ($\tau$) versus 1/T plot (see the insets of the lower panel (b) in Fig. 7). The activation energy and relaxation time obtained from the straight-line fit are: $E_a^{LT} = (0.167 \pm 0.004)$ eV, $\tau_0^{LT} = (2.1 \pm 0.6) \times 10^{-13}$ s and $E_a^{HT} = (0.203 \pm 0.002)$ eV, $\tau_0^{HT} = (7.5 \pm 2) \times 10^{-13}$ s, respectively, for the LT and HT dielectric relaxations. The observed value of $E_a$ is comparable to the values reported for polaronic relaxation in the range $E_a = 0.14 - 0.28$ eV caused by charge carrier hopping [49–53]. It is interesting to note that the observed peak temperatures nearly coincide with the magnetic transition temperatures suggesting that the charge and spin degrees of freedom are coupled and hint towards the possibility of magneto-dielectric coupling in SCRO. However, further field-dependent experiments would be required to confirm the magneto-dielectric effect in SCRO.

**Conclusions:**

In summary, we have presented temperature-dependent structure, magnetic, dielectric and transport properties of the ordered triple perovskite system $Sr_3CaRu_2O_9$. Stabilizing in the monoclinic *$P2_1/c$* symmetry at room temperature. We observe successive magnetic phase transitions ~190 K and ~160 K in the dc susceptibility studies. A clear lambda-like anomaly at 190 K in specific heat confirms the true thermodynamic nature of the para-antiferromagnetic phase transition. Temperature-dependent neutron powder diffraction measurements rule out within instrumental resolution, the possibility of a structural transition, and the system remains in the monoclinic *$P2_1/c$* symmetry down to 1.5 K. Interestingly, the integrated intensity of the magnetic peak ($I_{mag}$) decreases continuously with increasing temperature and vanishes around 200 K. We have modelled the temperature variation of $I_{mag}$ using the power law type functional dependence $I(T) \propto (1-T/T_N)^{2\beta}$ with where is critical exponent $\beta = (0.232 \pm 0.006)$ and $T_N = (201.3 \pm 0.2)$ K, respectively. Our results show that the unit cell volume, direct Ru-Ru distance, bond angles and dielectric permittivity/loss tangent change significantly across the magnetic transitions suggesting the presence of magnetoelastic and magnetodielectric couplings, respectively.

**Acknowledgments:**

Arun Kumar acknowledges SERB-DST, Government of India for providing financial support through a National Post-Doctoral Fellowship (PDF/2020/002116). Authors acknowledge Mr.

Sarath Kumar for synthesis of the $Sr_3CaRu_2O_9$ samples. SN acknowledges support from an Air Force Research Laboratory Grant (FA2386-21-1-4051).

**Figure Captions**

**Figure 1**: Panel (a) shows the observed (filled red circles), calculated (continuous black line) and difference (bottom blue line) profiles obtained after Rietveld refinement of X-ray powder diffraction data using the monoclinic $P2_1/c$ space group at 300 K for $Sr_3CaRu_2O_9$. The vertical tick marks (pink colour) indicate the expected Bragg peak positions. Panel (b) depicts the schematic crystal structure of $Sr_3CaRu_2O_9$ where violet, dark green, dark yellow and red colour balls represent the $Sr^{2+}$, $Ca^{2+}$, $Ru^{5+}$ and $O^{2-}$, respectively. It can be viewed as a 1:2 ordered structure with stacking sequence of one layer of $CaO_6$ octahedra and two layers of $RuO_6$ octahedra along the b-direction. Panel (c) represents a buckled honeycomb lattice formed by the Ru-atoms.

**Figure 2:** Upper panel (a): Temperature-dependent dc susceptibility of $Sr_3CaRu_2O_9$ measured at 1000 Oe applied magnetic field under zero-field cooled protocol. Middle panel (b): Temperature-dependence of specific heat at zero-magnetic field. Lower panel (c): Temperature dependence of real part of dielectric permittivity ($\varepsilon'$) measured at 100 kHz frequency. The upper inset of panel (c) depicts the magnified view of $\varepsilon'$ near the magnetic transition while the bottom inset shows the temperature dependence of loss tangent (tan$\delta$) to clearly discern the anomaly across the long-range ordered antiferromagnetic transition.

**Figure 3:** Panel (a) depicts the representative thermal evolution of neutron powder diffractogram of $Sr_3CaRu_2O_9$ recorded between 1.5 K to 250 K for 58° detector bank of WISH neutron diffractometer. Panel (b) shows the temperature evolution of the strongest magnetic peak after subtraction of the 220 K pattern as a paramagnetic background. This clearly indicates the appearance of a magnetic peak around 200 K. Panel (c) Variation of the integrated intensity of the strongest magnetic peak as a function of temperature. Solid line through data points represents the power law fit using the expression: $I(T) \propto (1 - T/T_N)^{2\beta}$ with $T_N = (201.3 \pm 0.2)$ and $\beta = (0.232 \pm 0.006)$ which is close to that reported in the tricritical system.

**Figure 4:** Observed (filled circles), calculated (continuous line), and difference (bottom line) profiles obtained from Rietveld refinement using *P2$_1$/c* space group at (a) 250 K and (b) 1.5 K. The vertical tick marks correspond to the position of all allowed Bragg reflections for the nuclear (top) and magnetic (bottom) reflections. Right panel shows the magnetic structure using Irrep $\Gamma_2$. The Ru$_1$ and Ru$_2$ spins are shown with black and blue colours, respectively.

**Figure 5:** Temperature dependence of (a) unit cell volume V, (b) direct Ru-Ru distance, and (c) Ru-O-Ru bond angles, as obtained by Rietveld refinement of NPD data of Sr$_3$CaRu$_2$O$_9$.

**Figure 6:** Main panel shows the temperature dependence of dc resistivity of Sr$_3$CaRu$_2$O$_9$. Inset: depict the plots corresponding to (i) simple Arrhenius model and (ii) Mott's variable-range hopping model. The solid lines through data points represent the fit for Eq. (2) and Eq. (3), respectively.

**Figure 7:** Temperature dependence of (a) real part of dielectric permittivity ($\varepsilon'$) and (b) loss tangent (tan$\delta$) at different frequencies for Sr$_3$CaRu$_2$O$_9$. Inset of panel (a) depicts the variation of dielectric permittivity on a magnified scale for LT transition. Insets of panel (b) show the ln ($\tau$) versus 1/T plots for LT and HT transition. Solid lines through data points represent the least-squares fit to the Arrhenius law as described in the text.

**Figures**

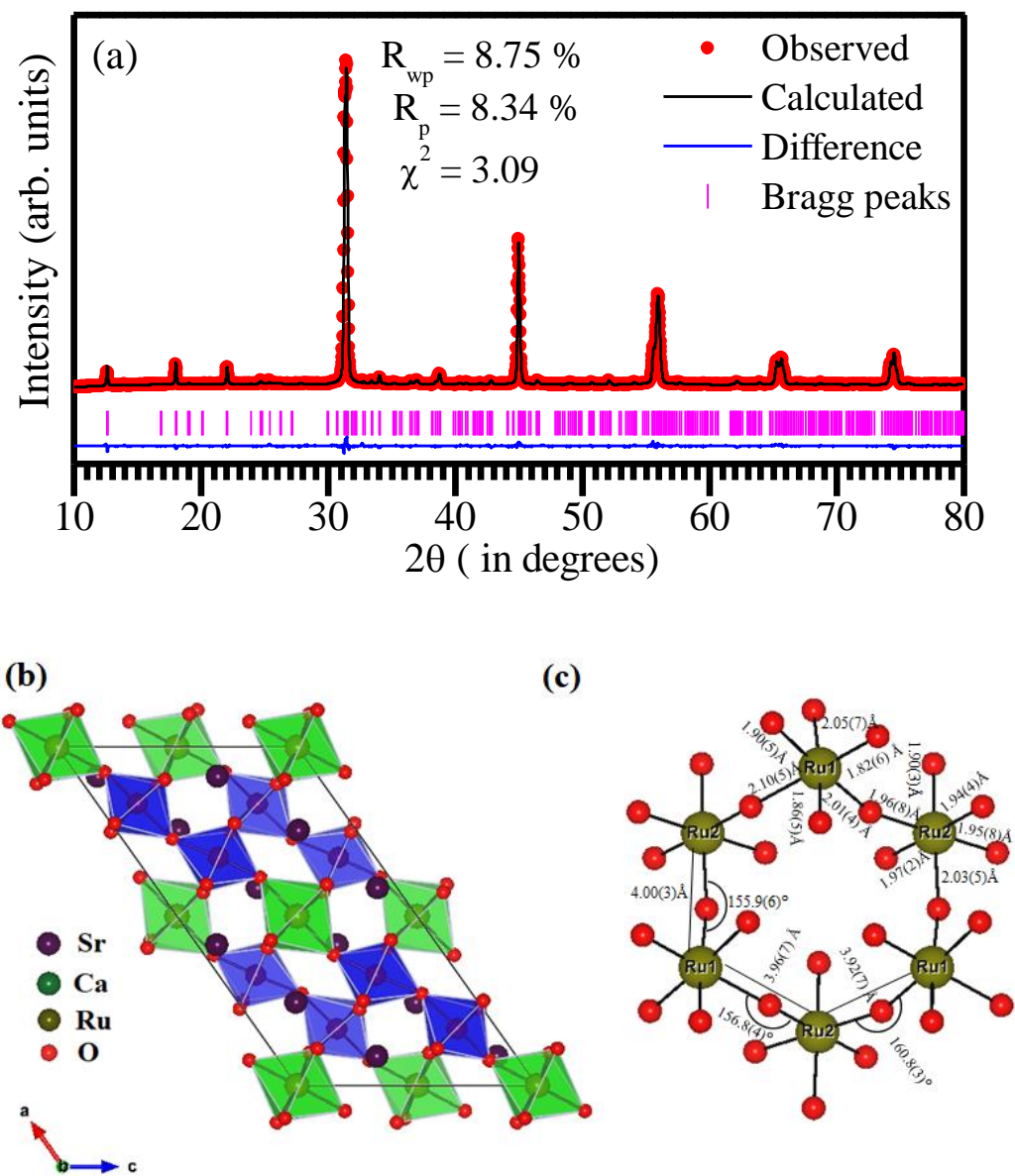

**Figure 1**

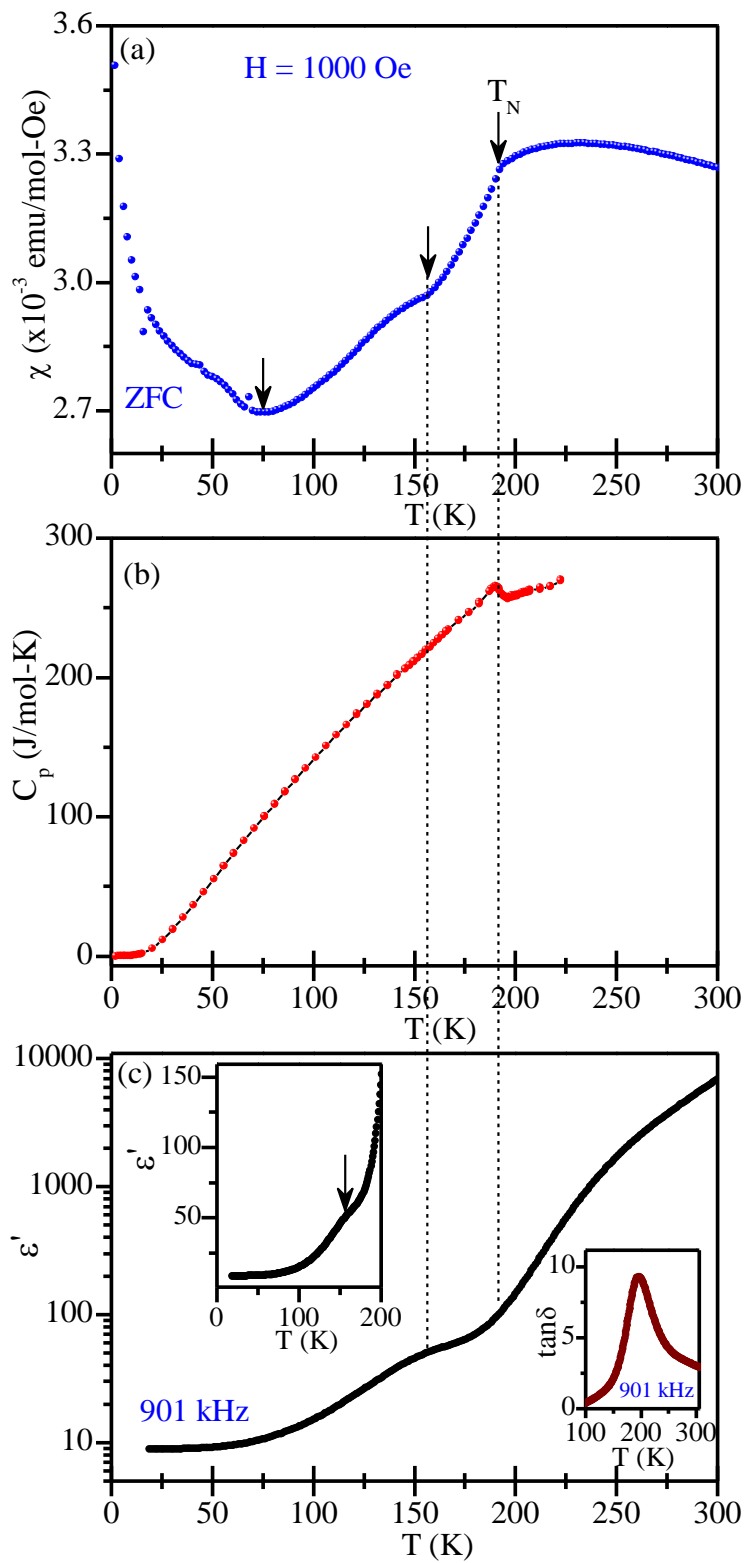

**Figure 2**

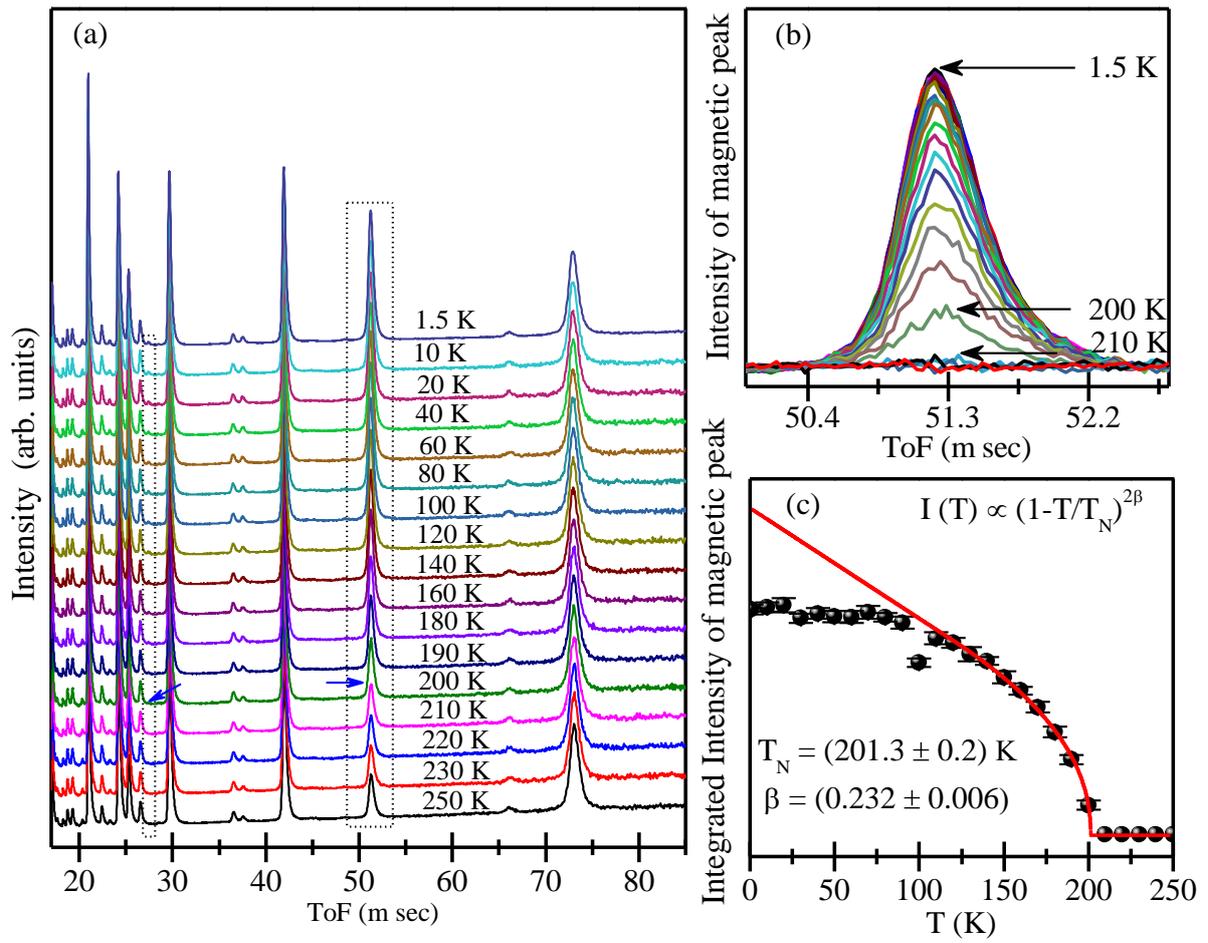

**Figure 3**

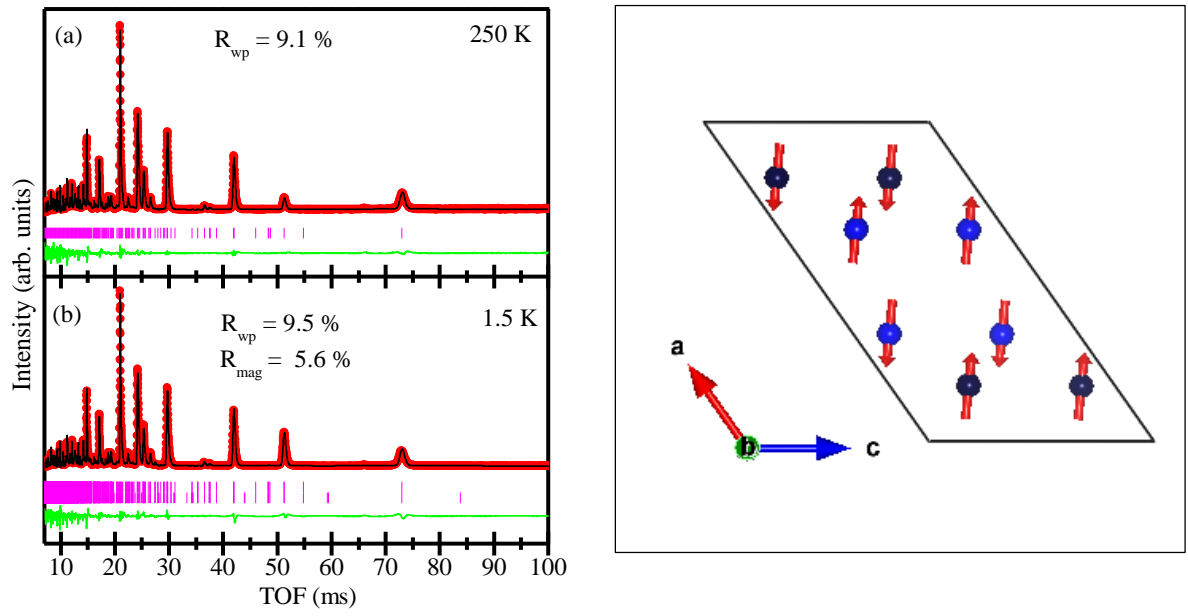

**Figure 4**

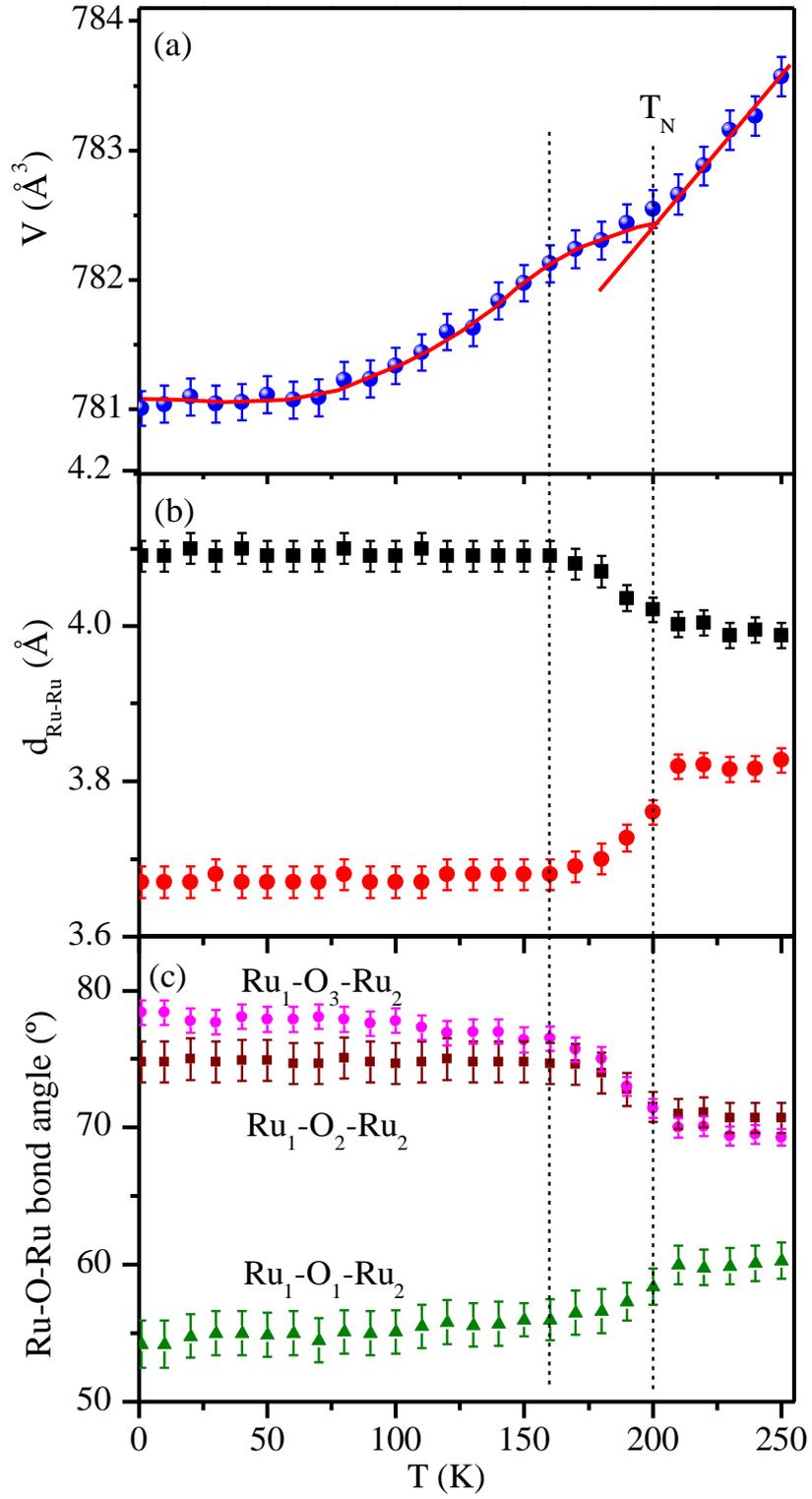

Figure 5

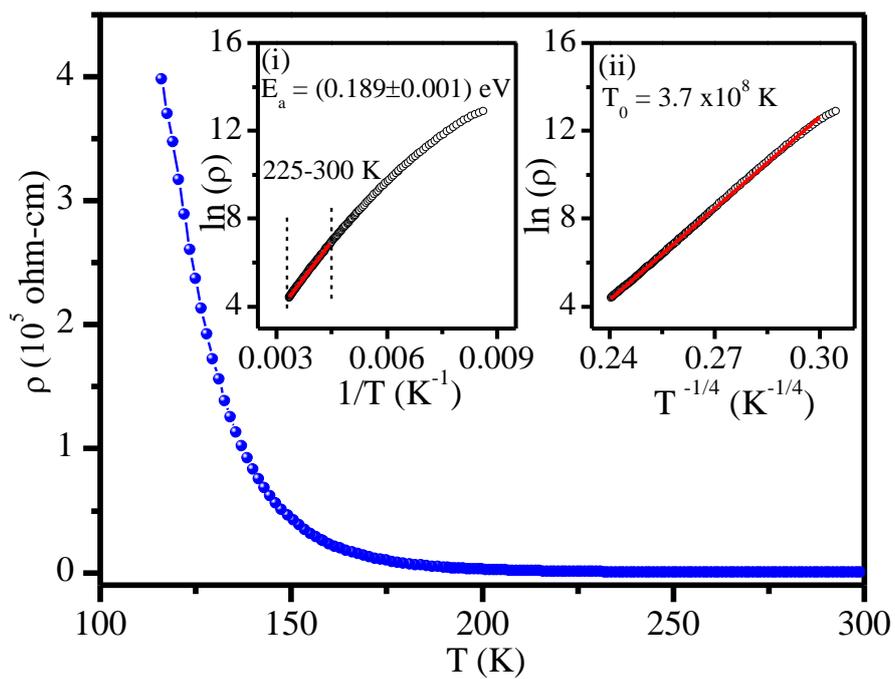

**Figure 6**

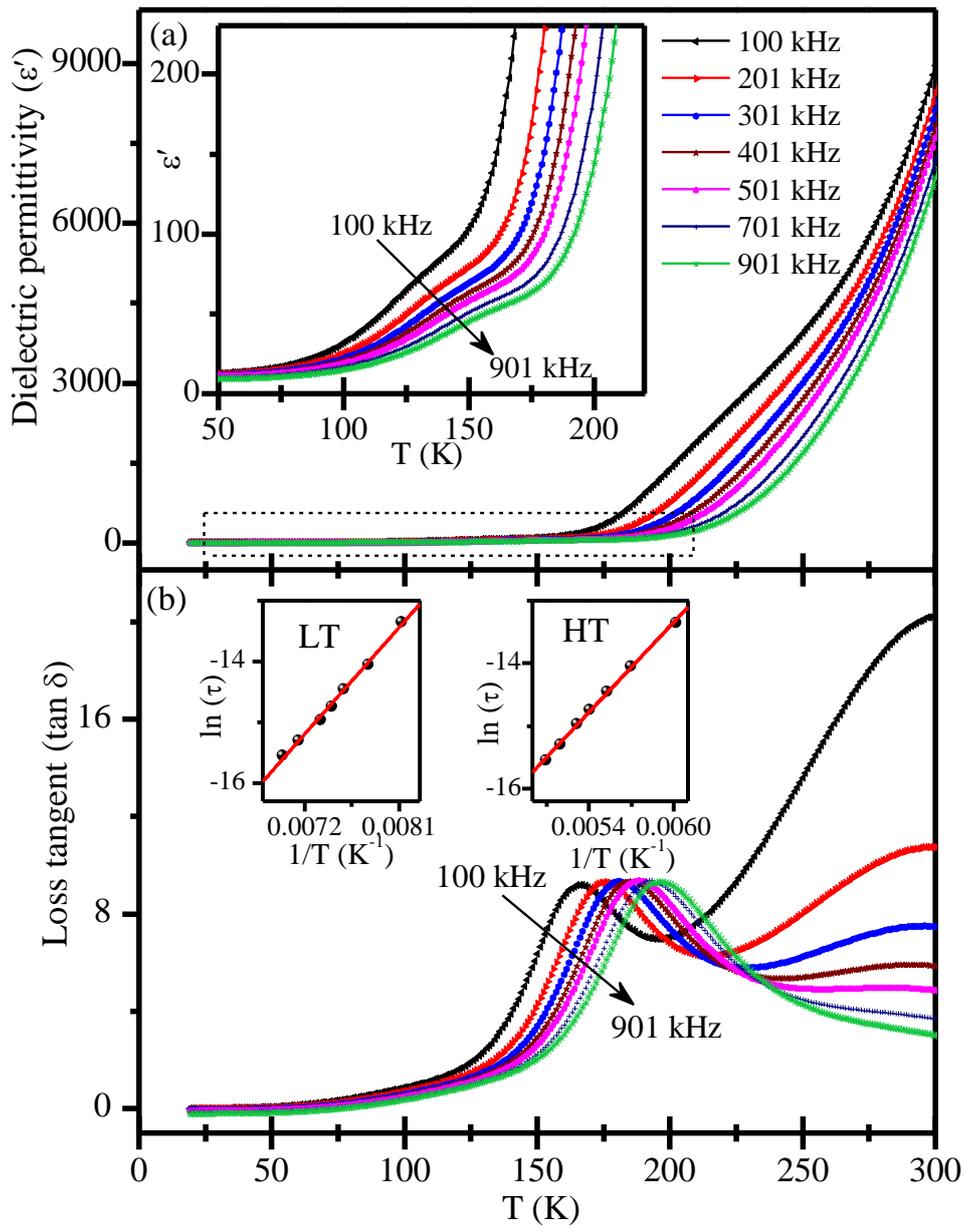

**Figure 7**

# Tables

**Table 1:** Structural parameters for Sr$_3$CaRu$_2$O$_9$ obtained from Rietveld refinement of XRD data using $P2_1/c$ space group.

| *a* (Å) | *b* (Å) | *c* (Å) | *β* (°) | *V* (Å$^3$) |
|---|---|---|---|---|
| 17.1433 (7) | 5.6836 (2) | 9.8485 (4) | 125.077 (0.003) | 785.321 (0.061) |
| *Atom* | *Wyckoff Site* | *x, y, z* | *Occ.* | *B$_{iso}$* (Å$^2$) |
| Sr1 | (4e) | 0.249(1), 0.519(1), 0.745(6) | 1 | 0.382 (5) |
| Sr2 | (4e) | 0.417(5), 0.032(4), 0.085(3) | 1 | 0.382 (5) |
| Sr3 | (4e) | 0.918(7), 0.513(5), 0.085(9) | 1 | 0.382 (5) |
| Ca1 | (2a) | 0, ½, ½ | 0.5 | 1.193 (9) |
| Ca2 | (2d) | ½, 0, ½ | 0.5 | 1.193 (9) |
| Ru1 | (4e) | 0.167(6), 0.007(5), 0.833(0) | 1 | 0.150 (1) |
| Ru2 | (4e) | 0.326(3), 0.502(6), 0.153(9) | 1 | 0.150 (1) |
| O1 | (4e) | 0.074(8), 0.491(6), 0.375(9) | 1 | 0.659 (7) |
| O2 | (4e) | 0.096(3), 0.802(5), 0.647(3) | 1 | 0.659 (7) |
| O3 | (4e) | 0.108(9), 0.257(4), 0.693(1) | 1 | 0.659 (7) |
| O4 | (4e) | 0.271(5), -0.019(3), 0.790(2) | 1 | 0.659 (7) |
| O5 | (4e) | 0.260(9), 0.211(3), 0.044(9) | 1 | 0.659 (7) |
| O6 | (4e) | 0.215(1), 0.793(7), 0.464(1) | 1 | 0.659 (7) |
| O7 | (4e) | 0.427(6), 0.744(1), 0.322(6) | 1 | 0.659 (7) |
| O8 | (4e) | 0.412(8), 0.517(8), 0.027(8) | 1 | 0.659 (7) |
| O9 | (4e) | 0.435(6), 0.230(4), 0.839(8) | 1 | 0.659 (7) |

**Table 2:** List of important bond lengths of $Sr_3CaRu_2O_9$ obtained from Rietveld refinement of XRD data.

| Atom | Oxygen | Mult. | Bond-length (Å) | Atom | Oxygen | Mult. | Bond-length (Å) |
|---|---|---|---|---|---|---|---|
| $Ca_1$ | $O_1$ | 2x | 2.2226 (3) | $Ca_2$ | $O_7$ | 2x | 2.2092 (9) |
|  | $O_2$ | 2x | 2.2433 (6) |  | $O_8$ | 2x | 2.2057 (2) |
|  | $O_3$ | 2x | 2.2186 (3) |  | $O_9$ | 2x | 2.2195 (8) |
| $Ru_1$ | $O_1$ | 1x | 1.8656 (3) | $Ru_2$ | $O_4$ | 1x | 2.0351 (3) |
|  | $O_2$ | 1x | 1.9044 (9) |  | $O_5$ | 1x | 1.9686 (0) |
|  | $O_3$ | 1x | 1.8260 (7) |  | $O_6$ | 1x | 1.9443 (8) |
|  | $O_4$ | 1x | 2.0578 (3) |  | $O_7$ | 1x | 1.9580 (6) |
|  | $O_5$ | 1x | 2.0144 (5) |  | $O_8$ | 1x | 1.9038 (9) |
|  | $O_6$ | 1x | 2.1054 (3) |  | $O_9$ | 1x | 1.9720 (1) |